\documentstyle[preprint,aps,floats,epsfig]{revtex}
\tightenlines

\newcommand{\ber}{\begin{eqnarray}}
\newcommand{\eer}{\end{eqnarray}}
\newcommand{\grpp}{g_{\rho \pi \pi}}
\newcommand{\gpdds}{g_{\pi D D^*}}
\newcommand{\grdd}{g_{\rho D D}}
\newcommand{\grdsds}{g_{\rho D^* D^*}}
\newcommand{\gprdds}{g_{\pi \rho D D^*}}
\newcommand{\grrr}{g_{\rho \rho \rho}}
\newcommand{\grrdd}{g_{\rho \rho D D}}
\newcommand{\gppdsds}{g_{\pi \pi D^* D^*}}
\newcommand{\grrdsds}{g_{\rho \rho D^* D^*}}
\newcommand{\ggp}{{\gamma^\prime}} 

\begin{document}
\title{Charm meson scattering cross sections by pion and rho meson}
\bigskip
\author{Ziwei Lin, T. G. Di, and C. M. Ko}
\address{Cyclotron Institute and Physics Department, Texas A\&M University,
College Station, Texas 77843-3366}
\maketitle

\begin{abstract}
Using the local flavor SU(4) gauge invariance in the limit of vanishing vector
meson masses, we extend our previous study of charm meson scattering
cross sections by pion and rho meson, which is based only on 
the pseudoscalar-pseudoscalar-vector meson couplings, 
to include also contributions from the couplings among three vector 
mesons and among four particles. We find that diagrams with light meson 
exchanges usually dominate the cross sections. 
For the processes considered previously, the additional interactions 
lead only to diagrams involving charm meson exchanges and contact interactions,
and the cross sections for these processes are thus not much affected. 
Nevertheless, these additional interactions introduce new processes with 
light meson exchanges and increase significantly the total scattering
cross sections of  charm mesons by pion and rho meson.

\medskip
\noindent PACS number(s): 25.75.-q, 13.75.Lb, 14.40.Lb

\end{abstract}

%%%%%%%%%%%%%%%%%%%%
\section{INTRODUCTION}

Since charm quarks may lose appreciable energies  
in a quark-gluon plasma via gluon radiations,
study of the charm meson spectrum in heavy ion 
collisions is expected to provide useful information on the properties of 
the quark-gluon plasma formed in these collisions \cite{closs1,closs2}.
However, charm mesons may interact strongly with hadrons during
later stage of heavy ion collisions, and this may also lead 
to changes in their final spectrum.  To use charm mesons as a 
probe of the properties of the initial quark-gluon plasma thus requires
the understanding of their interactions with hadrons.
In a previous study \cite{ds}, we have evaluated the charm meson scattering 
cross sections with pion and rho meson in a simple hadronic model that 
includes only the pseudoscalar-pseudoscalar-vector (PPV) meson interactions. 
In that study, we have also neglected the exchange of charm mesons 
as it is expected to be unimportant due to the large charm meson mass.
Including form factors at the interaction
vertices, we have obtained a thermally averaged total cross section 
of about 10 mb in the temperature range of interest. 
In a schematic model for the dynamics of 
heavy ion collisions, we have found that the inverse
slope of the charm meson transverse momentum spectrum is 
significantly enhanced by their scatterings in the hadronic matter.
As a result, the invariant mass spectrum of the dileptons from the
decay of charm meson pairs is expected to be modified, which
has been suggested as a possible explanation for the observed 
enhancement of intermediate-mass dileptons in heavy ion collisions 
at SPS energies \cite{dflow}. 

In Ref. \cite{muller}, Matinyan and M\"uller have used 
a similar hadronic Lagrangian 
to evaluate the cross sections of charmonium absorption in hadronic matter.
In contrast with the charm meson scattering cross sections, 
they have obtained very small charmonium absorption cross sections.
The model has been extended in Refs. \cite{haglin,jpsi} by using 
the local flavor SU(4) gauge invariance to include also interactions 
among three vector mesons and among four particles. Because of
these additional interactions, the charmonium absorption cross sections
are increased by an order-of-magnitude. In this paper, we shall use
this extended hadronic Lagrangian to study the charm meson scattering
cross sections by pion and rho meson.

The paper is organized as follows.  In Sec. \ref{s-eff}, we introduce
the hadronic Lagrangian based on the local flavor SU(4) gauge symmetry.
The interaction Lagrangians that are relevant 
to charm meson scattering with pion and rho meson are then given 
in Sec. \ref{s-ampl}. We also derive in this section
the scattering amplitudes for these processes and give their explicit 
expressions in Appendix A. Constraints on the scattering amplitudes
as a result of the conservation of SU(4) flavor current 
are then discussed in Sec. \ref{s-curr}, and an example is shown in detail 
in Appendix B. After addressing in Sec. \ref{s-div} 
the problem of on-shell divergence in some of the amplitudes, 
we fix the coupling constants in Sec. \ref{s-coup} and 
introduce in Sec. \ref{s-form} the form factors at interaction vertices. 
Numerical results for the charm meson scattering cross sections are presented 
in Sec. \ref{s-nume}. 
In Sec. \ref{s-comp}, we compare our results with previous ones
obtained using the PPV coupling and including only diagrams with
light meson exchanges. Finally, a summary is given in Sec. \ref{s-summ}.

%%%%%%%%%%%%%%%%%%%%
\section{CHARM MESON INTERACTIONS WITH HADRONS}

%%%%%%%%%%%%%%%%%%%%
\subsection{hadronic Lagrangian with SU(4) Symmetry}
\label{s-eff}

We have previously introduced a hadronic Lagrangian with SU(4) symmetry
for studying the charmonium absorption cross sections by 
hadrons \cite{jpsi}. It starts  from the  
free Lagrangian for pseudoscalar and vector mesons,
\begin{eqnarray}
{\cal L}_0= {\rm Tr} \left ( \partial_\mu P^\dagger \partial^\mu P \right )
-\frac{1}{2} {\rm Tr} \left ( F^\dagger_{\mu \nu} F^{\mu \nu} \right )~,
\label{lagn0}
\end{eqnarray}
where $F_{\mu \nu}=\partial_\mu V_\nu-\partial_\nu V_\mu$, 
and $P$ and $V$ denote, respectively, the $4\times 4$ pseudoscalar 
and vector meson matrices in SU(4) \cite{ds,jpsi}. 
Introducing the minimal substitution,
\begin{eqnarray}
\partial_\mu P &\rightarrow& {\cal D}_\mu P= \partial_\mu P
-\frac{ig}{2} \left [V_\mu, P \right ]~, \label{ms1} \\
F_{\mu \nu} &\rightarrow& 
\partial_\mu V_\nu-\partial_\nu V_\mu -\frac{ig}{2} 
\left [ V_\mu, V_\nu \right ]~, 
\label{ms}
\end{eqnarray}
leads to the following Lagrangian for the interacting hadrons:
\begin{eqnarray}
{\cal L}&=& {\cal L}_0 + ig {\rm Tr} 
\left ( \partial^\mu P \left [P, V_\mu \right ] \right ) 
-\frac{g^2}{4} {\rm Tr} 
\left ( \left [ P, V_\mu \right ]^2 \right ) \nonumber \\
&+& ig {\rm Tr} \left ( \partial^\mu V^\nu \left [V_\mu, V_\nu \right ] 
\right ) 
+\frac{g^2}{8} {\rm Tr} \left ( \left [V_\mu, V_\nu \right ]^2 \right )~.
\label{lagn2}
\end{eqnarray}
Since hadron masses explicitly break the SU(4) symmetry,
mass terms based on the experimentally determined values
are added to Eq. (\ref{lagn2}).

%%%%%%%%%%%%%%%%%%%%
\subsection{scattering amplitudes}
\label{s-ampl}

\begin{figure}[htp]
\centerline{\epsfig{file=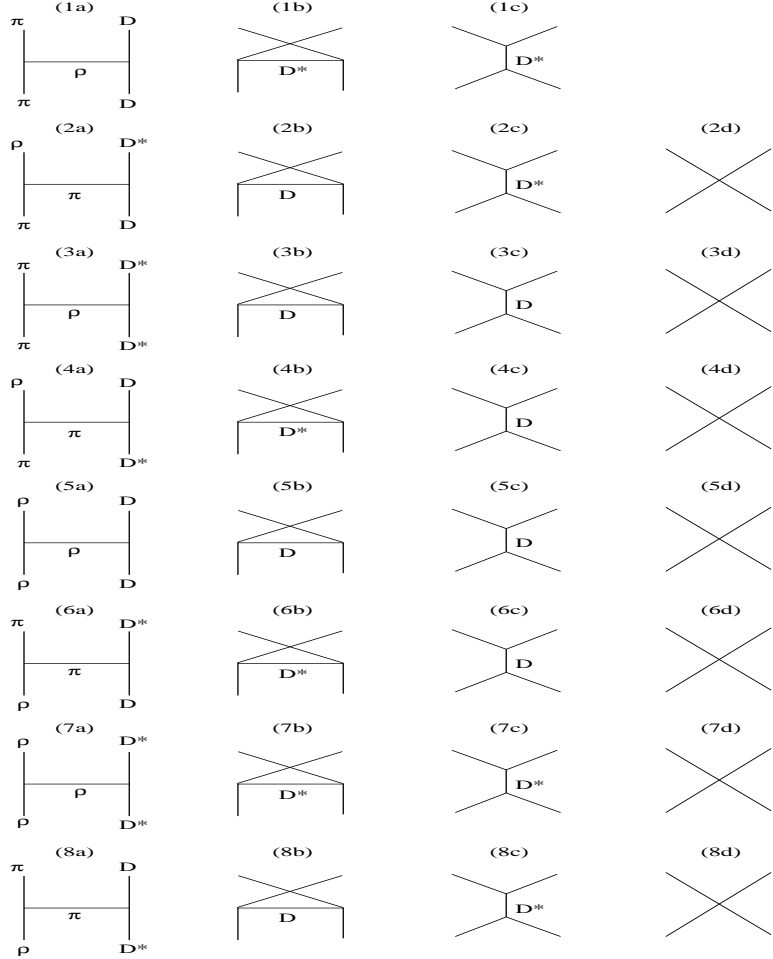,height=5in,width=4in,angle=0}}
\vspace{1cm}
\caption{Diagrams for charm meson scattering by pion and rho meson.   
The numbers denote different processes, and the Roman letters 
label the different amplitudes in a given process.}
\label{diagrams}
\end{figure}

The above Lagrangian yields the following processes 
for charm meson scattering by $\pi$ and $\rho$ mesons,
\begin{eqnarray}
\pi D \leftrightarrow \rho D^*,~
\pi D \rightarrow \pi D,~
\pi D^* \rightarrow \pi D^*,~
\pi D^* \leftrightarrow \rho D,~ 
\rho D \rightarrow \rho D,~
\rho D^* \rightarrow \rho D^*. 
\label{eight}
\end{eqnarray}
There are also similar processes for anti-charm mesons. 
Fig. \ref{diagrams} shows the diagrams for the eight processes 
in Eq. (\ref{eight}). 

Expanding the Lagrangian in Eq. (\ref{lagn2}) using the 
$4 \times 4$ matrices for $P$ and $V$, we obtain the following
interaction Lagrangians that are relevant to charm meson scattering:
\begin{eqnarray}
{\cal L}_{\rho \pi\pi}&=&g_{\rho \pi \pi} \vec
\rho^\mu \cdot \left ( \vec \pi \times \partial_\mu \vec \pi
\right ) \; ,\nonumber \\
{\cal L}_{\rho \rho\rho}&=&-g_{\rho \rho \rho} 
\partial_\mu \vec {\rho_\nu} \cdot \left ( \vec {\rho^\mu} \times 
\vec {\rho^\nu} \right ) \; ,\nonumber \\
{\cal L}_{\pi DD^*}&=&ig_{\pi DD^*}~ 
D^{* \mu} \vec \tau \cdot \left ( \bar D \partial_\mu \vec \pi - 
\partial_\mu \bar D \vec \pi \right ) + {\rm H.c.}~ , \nonumber \\ 
{\cal L}_{\rho DD}&=&ig_{\rho DD}~ \left ( D \vec \tau \partial_\mu \bar D
-\partial_\mu D \vec \tau \bar D \right ) \cdot \vec \rho^\mu ~ ,\nonumber \\ 
{\cal L}_{\rho D^*D^*}&=&ig_{\rho D^*D^*}~ \left [ 
\left ( \partial_\mu D^{* \nu} \vec \tau {\bar D}^*_\nu
-D^{* \nu} \vec \tau \partial_\mu {\bar D}^*_\nu \right ) \cdot \vec \rho^\mu 
\right .\nonumber \\ 
&+& \left . \left ( D^{* \nu} \vec \tau \cdot \partial_\mu \vec \rho_\nu 
-\partial_\mu D^{* \nu} \vec \tau \cdot \vec \rho_\nu \right ){\bar D}^{* \mu} 
+ D^{* \mu} \left ( \vec \tau \cdot \vec \rho^\nu \partial_\mu {\bar D}^*_\nu
-\vec \tau \cdot \partial_\mu \vec \rho^\nu {\bar D}^*_\nu \right ) \right ] 
~ ,\nonumber \\ 
{\cal L}_{\pi\rho DD^*}&=&-g_{\pi\rho DD^*}~  
D^{*\mu} \left ( 2\vec \tau \cdot \vec \pi \vec \tau \cdot \vec {\rho_\mu}
-\vec \tau \cdot \vec {\rho_\mu} \vec \tau \cdot \vec \pi \right ) \bar D
+{\rm H.c.}~ ,\nonumber \\
{\cal L}_{\pi\pi D^* D^*}&=&g_{\pi\pi D^* D^*}
\left ( \frac { \vec \pi \cdot \vec \pi} {2} \right ) D^{* \mu} {\bar D}^*_\mu 
,\nonumber \\
{\cal L}_{\rho\rho DD}&=&g_{\rho \rho DD}~  
\left (\frac { \vec {\rho_\mu} \cdot \vec {\rho^\mu}} {2} \right ) D \bar D 
,\nonumber \\
{\cal L}_{\rho\rho D^*D^*}&=&g_{\rho \rho D^*D^*}~  D^{*\mu} \left ( 
2\vec \tau \cdot \vec {\rho_\nu} \vec \tau \cdot \vec {\rho_\mu} 
-\vec \tau \cdot \vec {\rho_\mu} \vec \tau \cdot \vec {\rho_\nu} 
-\vec {\rho_\gamma} \cdot \vec {\rho^\gamma} g_{\mu \nu}
\right ) {\bar D}^{*\nu}.
\end{eqnarray}
In the above, $\vec \tau$ are Pauli matrices;
$\vec \pi$ and $\vec \rho$ denote the pion and rho meson isospin triplets, 
respectively; while $D$ and $D^*$ denote the pseudoscalar and vector
charm meson isospin doublets, respectively.  

Using the above interacting Lagrangians, we have derived
the amplitudes for all diagrams 
in Fig. \ref{diagrams}, and they are given 
in Appendix A. In general, the amplitude for a process $n$, 
before summing and averaging over external spins and isospins,  
is given by the coherent sum of all individual amplitudes that 
contributing to the process, i.e., 
\begin{eqnarray}\label{reduce} 
{\cal M}_n 
=\left ( \sum_{i} {\cal M}_{n i}^{\lambda_k \cdots \lambda_l} \right )
\epsilon_{k \lambda_k} \cdots \epsilon_{l \lambda_l} 
\equiv {\cal M}_n^{\lambda_k \cdots \lambda_l}~
\epsilon_{k \lambda_k} \cdots \epsilon_{l \lambda_l} 
\end{eqnarray}
where $i$ runs through 
$a,b,c$ for process 1 and $a,b,c,d$ for all other processes, 
and $\epsilon_{j\lambda_j}$ denotes the polarization vector of external 
vector meson $j$. 

%%%%%%%%%%%%%%%%%%%%
\subsection{current conservation}
\label{s-curr}

Since the Lagrangian in Eq. (\ref{lagn2}) is generated from the free 
Lagrangian in Eq. (\ref{lagn0}) by the minimal substitution, it is 
invariant under the local flavor SU(4) gauge transformation, 
thus it is also invariant under the global flavor SU(4) gauge transformation. 
This invariance
remains valid after including degenerate pseudoscalar and degenerate
vector meson mass terms, leading to the conservation of a SU(4) flavor
current. As a result, the scattering amplitude for any process 
satisfies the following condition: 
\begin{eqnarray}\label{conservation} 
{\cal M}_n^{\lambda_k \dots \lambda_l}~ p_{j \lambda_j}=0~,
\end{eqnarray}
where $p_{j\lambda_j}$ is the momentum of external vector meson $j$.
As an example, we show explicitly in Appendix B that the condition 
${\cal M}_2^{\lambda \omega} p_{4 \omega}=0$ is indeed satisfied by
the amplitude for process 2, $\pi D\to\rho D^*$.

%%%%%%%%%%%%%%%%%%%%
\subsection{on-shell divergence}
\label{s-div}

The amplitudes for diagrams 3b, 4a, and 6a become singular 
when the exchanged mesons are on-shell. Since the on-shell process 
describes a two-step process, their contribution needs to be subtracted 
from the cross section. Several methods have 
been proposed to treat such a singularity \cite{singular}. 
Since we are interested in charm meson scattering in hadronic
matter, the exchanged meson is expected to acquire an imaginary 
self-energy due to collisional broadening. The one-step process then
corresponds to keeping only the real part of the propagator for the 
exchanged meson. However, a consistent evaluation of this effect 
also requires the inclusion of vertex corrections due to the medium, 
which has not been carried out even for light meson scattering in 
hadronic matter.  
We thus follow Ref. \cite{ds} by adding an imaginary part of 50 MeV to the 
self-energy of the exchanged meson in the above three diagrams. 
Since the width of $D^*$ in vacuum is very small (about 44 KeV) \cite{dwidth}, 
the amplitude for diagram 1c can also be very large when the center-of-mass
energy of the initial pion and charm meson is close to the $D^*$ mass. 
We thus also add an imaginary part of 50 MeV to the self-energy of 
the $D^*$ meson in diagram 1c. In Sec. \ref{s-nume} we shall show that 
thermal averages of these cross sections do not change much for 
values of imaginary self-energy between $5$ and $500$ MeV. 

%%%%%%%%%%%%%%%%%%%%
\subsection{coupling constants}
\label{s-coup}

For the coupling constants in the interaction Lagrangians, 
we shall use empirical values if they are available, i.e. 
$g_{\rho \pi \pi}=6.1$\cite{rpp}, 
$g_{\pi DD^*}=4.4$ \cite{dwidth}.
$g_{\rho DD}=g_{\rho D^* D^*}=2.52$ \cite{muller,jpsi}.
Since there is little empirical information on
other coupling constants, 
we use the SU(4) relations to determine their values, i.e., 
\begin{eqnarray}
g_{\rho \rho \rho}= g_{\rho \pi \pi}, ~
g_{\pi \rho DD^*}= g_{\pi DD^*} g_{\rho DD}, ~
g_{\pi \pi D^* D^*}= 2~g_{\pi DD^*}^2, ~
g_{\rho \rho D D}= 2~g_{\rho DD}^2, ~
g_{\rho \rho D^* D^*}=g_{\rho D^* D^*}^2.  
\end{eqnarray}

We note that the SU(4) symmetry gives the following relations among
couplings constants: 
\begin{eqnarray} 
&&\frac {g_{\rho \pi \pi}}{2} (3.0)
=\gpdds (4.4)
=g_{\rho DD} (2.5)
=\frac {g_{\rho \rho \rho}}{2}
=\grdsds = \frac{g}{4} ; \nonumber \\
&&g_{\pi \rho DD^*}
=\frac {g_{\pi \pi D^* D^*}}{2}
=\frac {g_{\rho \rho D D}}{2}
=g_{\rho \rho D^* D^*}= \frac{g^2}{16}. 
\label{coupling}
\end{eqnarray}
The empirical values given in the parentheses are seen to 
agree reasonably with those predicted by the SU(4) symmetry. 

%%%%%%%%%%%%%%%%%%%%
\subsection{form factors}
\label{s-form}

Because of the finite size of hadrons, form factors are needed at  
interaction vertices. In the present study, 
we take them to be the usual mono-pole form 
for vertices in the $t$ and $u$ channel processes, i.e., 
\begin{eqnarray}
f_3(t~ {\rm or}~ u)=\frac {\Lambda^2}{\Lambda^2+{\bf q}^2},
\label{ff1}
\end{eqnarray}
where $\Lambda$ is a cutoff parameter, and 
${\bf q}^2$ is the squared three momentum transfer in the 
center-of-mass frame, given by $({\bf p_1}-{\bf p_3})_{\rm cm}^2$   
and $({\bf p_1}-{\bf p_4})_{\rm cm}^2$, respectively, 
for  the $t$ and $u$ channel processes. These form factors are different from
that used in Ref. \cite{ds}, where it is given by 
$f(t)=(\Lambda^2-m^2)/(\Lambda^2-t)$, since the latter is not suitable
for diagrams involving the charmed meson exchange that
has a large invariant four momentum transfer $t$. 

As in Ref. \cite{ffs}, form factors at $s$ channel vertices are taken as
\begin{eqnarray}
f_3(s)=\frac {\Lambda^2}{\Lambda^2+{\bf k}^2},
\label{ff2}
\end{eqnarray}
with ${\bf k}$ denoting the three momentum of either the incoming or 
outgoing particles in the center-of-mass, i.e., 
${\bf k}^2=p_{i,\rm cm}^2$ or $p_{f,\rm cm}^2$.    

After introducing the form factors at three-point vertices, the 
form factors at four-point vertices can in principle be determined 
by requiring the total amplitude for a given process satisfies the
current conservation condition of Eq. (\ref{conservation})
\cite{ff,ffj}.  Since the uncertainty of form factors involving 
charm mesons is already large for three-point vertices and the gauge 
invariance is not valid once we use empirical vector meson masses, 
we choose not to follow this more involved approach. Instead, we simply 
take the form factors at four-point vertices to be
\begin{eqnarray}
f_4=\left ( \frac {\Lambda^2}{\Lambda^2+\bar {{\bf q}^2} } \right )^2, 
\label{ff3}
\end{eqnarray}
where $\bar {{\bf q}^2}$ is the average value 
of the squared three momenta in the form factors 
for the $s$, $t$, and $u$ channels, i.e., 
\begin{eqnarray}
\bar {{\bf q}^2}=\frac {5}{6} \left ({p_{i,\rm cm}^2+p_{f,\rm cm}^2} \right ).
\end{eqnarray}

Since there is no empirical information on form factors involving charm
mesons, we shall use for simplicity the same value for all cutoff parameters
and choose $\Lambda$ as either $1$ or $2$ GeV to study the uncertainties
of our results due to form factors.

%%%%%%%%%%%%%%%%%%%%
\subsection{thermally averaged cross sections}
\label{s-sv}

For charm meson scattering in hadronic matter, it is useful to 
study the thermal average of their cross sections. For 
a hadronic matter at temperature $T$, this is given by 
\begin{eqnarray} 
\langle \sigma v \rangle 
&=&\frac{\int_{z_0}^{\infty} dz \left [z^2-(\alpha_1+\alpha_2)^2
\right ] \left [z^2-(\alpha_1-\alpha_2)^2 \right ] K_1(z) ~
\sigma (s=z^2{\rm T}^2)}
{4 \alpha_1^2 K_2(\alpha_1)\alpha_2^2 K_2(\alpha_2)} ~,  
\end{eqnarray} 
where $\alpha_i=m_i/{\rm T}$, 
$z_0={\rm max}(\alpha_1+\alpha_2,\alpha_3+\alpha_4)$, 
$K_n$'s are modified Bessel functions, and 
$v$ is the relative velocity of initial-state 
particles in their collinear frame \cite{err}. 

%%%%%%%%%%%%%%%%%%%%
\section{NUMERICAL RESULTS}
\label{s-nume}

\begin{figure}[htp]
\centerline{\epsfig{file=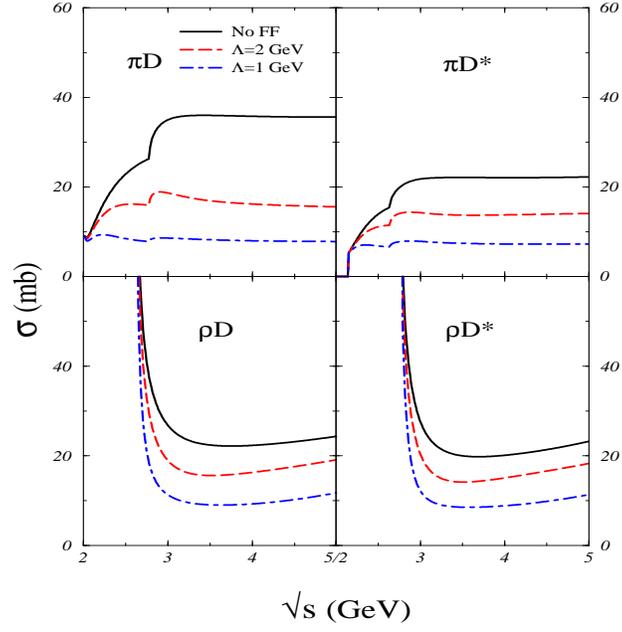,height=3.3in,width=3.3in,angle=0}}
\vspace{1cm}
\caption{Total cross section as functions of energy 
without and with form factors.}
\label{s-3ff}
\end{figure}

\begin{figure}[htp]
\centerline{\epsfig{file=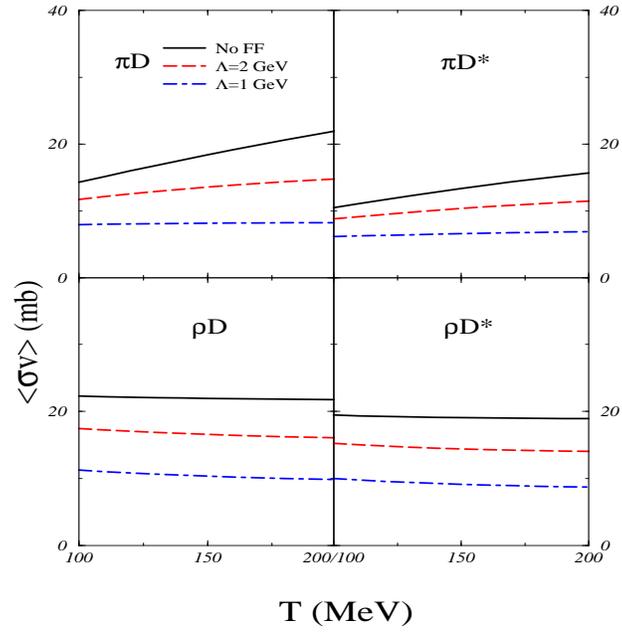,height=3.3in,width=3.3in,angle=0}}
\vspace{1cm}
\caption{Thermal average of the total cross section as functions of 
temperature without and with form factors.}
\label{sv-3ff}
\end{figure}

We first consider the case without form factors at interaction vertices, 
i.e., $\Lambda=\infty$. In Fig. \ref{s-3ff}, the solid lines 
show the energy dependence of the total 
charm meson scattering cross section for a given initial state, i.e.,
processes 1 and 2 for $\pi D$ scattering,  
processes 3 and 4 for $\pi D^*$ scattering,  
processes 5 and 6 for $\rho D$ scattering,  
and processes 7 and 8 for $\rho D^*$ scattering.   
We have also evaluated the thermal average of these cross sections,
and they are shown by the solid lines in 
Fig. \ref{sv-3ff} as functions of temperature.

To study the effects due to form factors, 
we take the value for the cutoff parameter $\Lambda$ as either $2$ or $1$ GeV.
The results are shown in Figs. \ref{s-3ff} and \ref{sv-3ff} 
by the dashed and dash-dotted curves, respectively.  
As expected, magnitude of the cross sections decreases with decreasing 
cutoff parameter. For the cutoff parameters used here, the 
cross sections for $\pi D, \pi D^*, \rho D$ and $\rho D^*$ scatterings
are all roughly between 10 and 20 mb. Compared to the case without form 
factors, we see that form factors only suppress modestly 
(by a factor of two of less) the total cross sections 
and their thermal averages. This is due to the dominance of
elastic processes, which involve small momentum transfer near the 
threshold. In contrast, form factors suppress significantly
the cross section for $J/\psi$ absorption by pion, e.g.,   
the process $\pi \psi \rightarrow D^* \bar D$ is reduced 
by as much as a factor of 8 due to its large threshold \cite{jpsi}.

\begin{figure}[htp]
\centerline{\epsfig{file=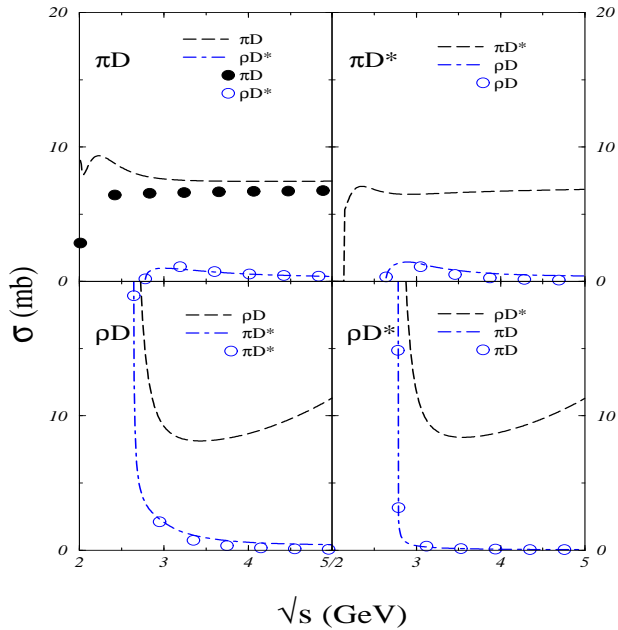,height=3.3in,width=3.3in,angle=0}}
\vspace{1cm}
\caption{Individual cross sections with form factors ($\Lambda=1$ GeV)
as functions of energy. Circles represent results based on the
previous study that includes only the pseudoscalar-pseudoscalar-vector
meson couplings and the light meson exchanges {\protect \cite{ds}}.}  
\label{s-ff1}
\end{figure}

\begin{figure}[htp]
\centerline{\epsfig{file=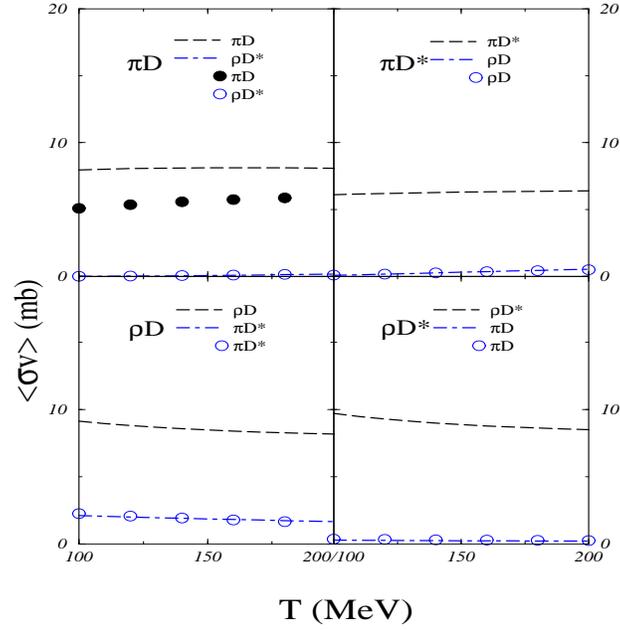,height=3.3in,width=3.3in,angle=0}}
\vspace{1cm}
\caption{Thermal averages of individual 
cross sections with form factors ($\Lambda=1$ GeV) as functions of 
temperature.  Circles represent results based on the
previous study that includes only the pseudoscalar-pseudoscalar-vector
meson couplings and the light meson exchanges {\protect \cite{ds}}.}
\label{sv-ff1}
\end{figure}

In Figs. \ref{s-ff1} and \ref{sv-ff1}, we show  
by the dashed and dot-dashed curves the
cross sections for individual processes and their thermal averages 
for a cutoff parameter of $1$ GeV.
It is seen that cross sections for both processes 2 
($\pi D \rightarrow \rho D^*$) and 4 ($\pi D^* \rightarrow \rho D$)
increase from zero 
at their respective threshold, while cross sections for processes 6 
($\rho D \rightarrow \pi D^*$) 
and 8 ($\rho D^* \rightarrow \pi D$) 
diverge near threshold because they are exothermic.
For the four elastic processes 1, 3, 5, and 7, 
their cross sections are finite at threshold. 
We also note that elastic processes are much more important than 
corresponding inelastic processes. 

\begin{figure}[htp]
\centerline{\epsfig{file=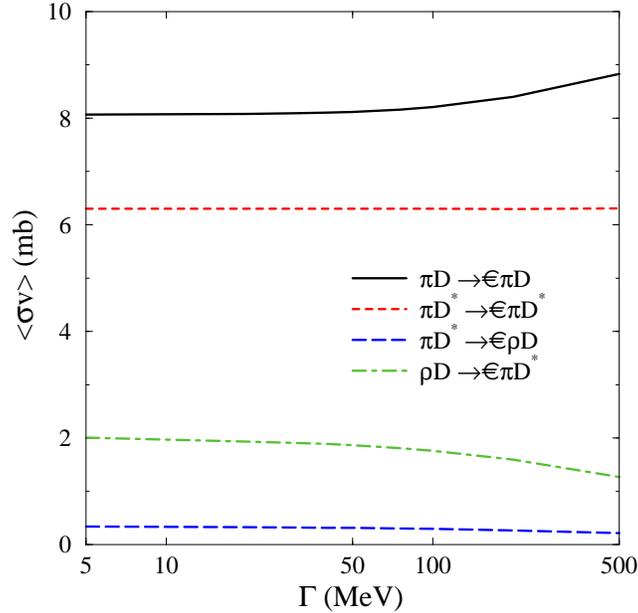,height=3.3in,width=3.3in,angle=0}}
\vspace{1cm}
\caption{Thermal average of cross sections at 
$T=150$ MeV with form factors ($\Lambda=1$ GeV) as functions of the 
imaginary self-energy of the exchanged meson.}
\label{sv-imag}
\end{figure}

Fig. \ref{sv-imag} shows the thermal averages of cross sections for 
processes 1, 3, 4 and 6 at $T=150$ MeV as functions of the imaginary 
self-energy $\Gamma$ of the exchanged meson. It is seen that 
they do not vary much for values of $\Gamma$ between 5 and 500 MeV.

%%%%%%%%%%%%%%%%%%%%
\section{Comparison with Previous Results}
\label{s-comp}

In our previous study of charm meson scattering cross sections \cite{ds},
we have considered only the pseudoscalar-pseudoscalar-vector meson 
interactions, which lead to diagrams
1a-c, 2a-b, 3b-c, 4a, 4c, 5b-c, 6a, 6c, and 8a-b in Fig. \ref{diagrams}. 
Since diagrams involving charm mesons are expected to be less important, 
we have evaluated only those diagrams that involve the exchange
of light mesons ($\pi$ or $\rho$ meson), i.e., diagrams (a)
in processes 1, 2, 4, 6, and 8. These results are shown
by circles in Figs. \ref{s-ff1} and \ref{sv-ff1}.
Except for process 1 ($\pi D \rightarrow \pi D$) near the threshold,
they are close to present results that include also
charm meson exchanges and contact terms. 
This comparison thus
demonstrates explicitly that diagrams with light meson exchanges
dominate the charm meson scattering cross sections.
This is in contrast to the charmonium absorption cross section in 
hadronic matter, where inclusion of additional couplings among 
three vector mesons and among four particles
increases the $J/\psi$ absorption cross section by pion 
by an order of magnitude \cite{haglin,jpsi}. 
This is due to the absence of light meson exchanges 
in charmonium absorption by hadrons.

We note that these additional interactions 
among three vector mesons and among four particles 
yield new processes with light meson exchanges, i.e., diagrams 3a, 5a, and 7a. 
We have checked that these diagrams also dominate 
the cross sections for these processes.
As shown in Fig. \ref{s-ff1} and \ref{sv-ff1},  
these elastic processes are more important than corresponding inelastic 
processes and thus increase significantly
the total scattering cross sections of charm mesons by pion and rho meson.

%%%%%%%%%%%%%%%%%%%%
\section{SUMMARY}
\label{s-summ}

In summary, we have studied the scattering cross sections of charm
mesons by pion and rho meson using a gauge invariant hadronic Lagrangian 
generated from the SU(4) symmetry. This leads to interaction Lagrangians
not only among two pseudoscalar mesons and one vector meson but also 
among three vector mesons as well as among four particles.
We have found that the charm meson scattering cross sections
are dominated by diagrams with light meson exchanges.
For the processes considered previously based only on 
the pseudoscalar-pseudoscalar-vector meson interactions, 
these additional interaction Lagrangians do not introduce 
new diagrams with light meson exchanges, and
their cross sections are thus not much affected. 
However, the interaction Lagrangians involving three vector 
mesons or four particles lead to new processes with light meson 
exchanges besides those considered previously and thus increase the 
total charm meson scattering cross sections by hadrons.
Therefore, we expect as in the previous study that the 
charm meson spectra in heavy ion collisions can be significantly modified
by hadronic scattering.

%%%%%%%%%%%%%%%%%%%%%%%%%%%%%%%%%%%%%%%%%%%%%%%%%%%%%%%%%%%%%%%%%%%%%
\section*{ACKNOWLEDGMENTS} 

This work was supported in part by the National Science Foundation under 
Grant No. PHY-9870038, the Welch Foundation under Grant No. A-1358,
and the Texas Advanced Research Program under Grant Nos. FY97-010366-0068
and FY99-010366-0081.

\pagebreak

%%%%%%%%%%%%%%%%%%%%
\section*{Appendix A}
\setcounter{equation}{0}
\def\theequation{A\arabic{equation}}

In this appendix, we give the explicit expressions for the amplitudes of
all diagrams in Fig. \ref{diagrams}. 
We show only the reduced amplitudes without the polarization
vectors of external vector mesons and
before summing and averaging over external spins and isospins.
 
For process 1, $\pi D \rightarrow \pi D$, we have
\ber
{\cal M}_{1a} &=& \grpp \grdd~ 
\left ( -\imath \epsilon_{ijk} \tau^k \right )_{\alpha \beta} 
\left ( \frac {1}{t-m_\rho^2} \right ) (s-u), \nonumber \\
{\cal M}_{1b} &=& \gpdds^2~ (\tau^j \tau^i)_{\alpha \beta} 
\left ( \frac {1}{u-m_{D^*}^2} \right ) 
\left [ s-t-\frac{(m_D^2-m_\pi^2)^2}{m_{D^*}^2} \right ], \nonumber \\
{\cal M}_{1c} &=& \gpdds^2~ (\tau^i \tau^j)_{\alpha \beta} 
\left ( \frac {1}{s-m_{D^*}^2} \right ) 
\left [ -t+u-\frac{(m_D^2-m_\pi^2)^2}{m_{D^*}^2} \right ].
\eer

For process $2$, $\pi D \rightarrow \rho D^*$, we have
\begin{eqnarray}
{\cal M}_{2a}^{\lambda \omega}&=& \grpp \gpdds~ 
\left ( \imath \epsilon_{ijk} \tau^k \right )_{\alpha \beta} 
(-2p_1+p_3)^\lambda \left ( \frac{1} {t-m_\pi^2} \right )
(p_1-p_2-p_3)^\omega, \nonumber \\
{\cal M}_{2b}^{\lambda \omega}&=& \gpdds \grdd~ (\tau^j \tau^i)_{\alpha \beta}
(-p_1+p_2+p_4)^\lambda \left ( \frac{1}{u-m_D^2} \right )
(-2p_1+p_4)_\omega, \nonumber \\
{\cal M}_{2c}^{\lambda \omega}&=& \gpdds \grdsds 
(\tau^i \tau^j)_{\alpha \beta}
(p_1-p_2)^\gamma  \left ( \frac{1}{s-m_{D^*}^2} \right )
\left [ g_{\gamma \ggp}-\frac{(p_1+p_2)_\gamma 
(p_1+p_2)_\ggp}{m_{D^*}^2} \right ] \nonumber \\
& \times &
\left [ (-p_1-p_2-p_4)^\lambda g^{\ggp \omega} + 
(-p_3+p_4)^\ggp g^{\lambda \omega} + 
(p_1+p_2+p_3)^\omega g^{\ggp \lambda} \right ] 
,\nonumber \\
{\cal M}_{2d}^{\lambda \omega}&=& \gprdds~ 
(\tau^i \tau^j-2\tau^j \tau^i)_{\alpha \beta} g^{\lambda \omega}.
\label{m2}
\end{eqnarray}

In the above, $p_n$ denotes the momentum of particle $n$.
Our convention is such that particles $1$ and $2$ represent 
initial-state mesons while particles $3$ and $4$ 
represent final-state mesons on the left and right side of the diagrams 
shown in Fig. \ref{diagrams}, respectively.
For vector mesons, the indices $\mu, \nu, \lambda$, and $\omega$ 
denote the polarization components of external mesons while
the indices $\gamma$ and $\ggp$ denote those of the exchanged meson. 
The indices $i$ and $j$ represent the isospin state of isospin-triplet 
mesons on the left of a diagram, while the indices $\alpha$ and $\beta$ 
represent those of isospin-doublet mesons on the right of a diagram. 
For the isospin-triplet meson in the propagator  
the index $k$ represents its isospin state.

The amplitudes for processes 4, 6 and 8 are related to above 
amplitudes for process 2 by the crossing symmetry. 
For process 4, $\pi D^* \rightarrow \rho D$, we then have 
\ber
{\cal M}_{4a}^{\nu \lambda}&=& 
{\rm \hat {T_4}} {\cal M}_{2a}^{\lambda \omega}, 
{\cal M}_{4b}^{\nu \lambda}= 
{\rm \hat {T_4}} {\cal M}_{2c}^{\lambda \omega}, 
{\cal M}_{4c}^{\nu \lambda}= 
{\rm \hat {T_4}} {\cal M}_{2b}^{\lambda \omega}, 
{\cal M}_{4d}^{\nu \lambda}= 
{\rm \hat {T_4}} {\cal M}_{2d}^{\lambda \omega},
\label{mi}
\eer
where ${\rm \hat {T_4}}$ represents the replacement, 
$p_2 \leftrightarrow -p_4, \nu \leftrightarrow \omega$, and 
$i \leftrightarrow j$.  
For process 6, $\rho D \rightarrow \pi D^*$, we have
\ber
{\cal M}_{6a}^{\mu \omega}&=& 
{\rm \hat {T_6}} {\cal M}_{2a}^{\lambda \omega}, 
{\cal M}_{6b}^{\mu \omega}= 
{\rm \hat {T_6}} {\cal M}_{2c}^{\lambda \omega}, 
{\cal M}_{6c}^{\mu \omega}= 
{\rm \hat {T_6}} {\cal M}_{2b}^{\lambda \omega}, 
{\cal M}_{6d}^{\mu \omega}= 
{\rm \hat {T_6}} {\cal M}_{2d}^{\lambda \omega},
\eer
where ${\rm \hat {T_6}}$ represents the replacement,  
$p_1 \leftrightarrow -p_3, \mu \leftrightarrow \lambda$, and 
$i \leftrightarrow j$. 
For process 8, $\rho D^* \rightarrow \pi D$, we have 
\ber
{\cal M}_{8a}^{\mu \nu}&=& 
{\rm \hat {T_{8}}} {\cal M}_{2a}^{\lambda \omega}, 
{\cal M}_{8b}^{\mu \nu}= 
{\rm \hat {T_{8}}} {\cal M}_{2b}^{\lambda \omega}, 
{\cal M}_{8c}^{\mu \nu}= 
{\rm \hat {T_{8}}} {\cal M}_{2c}^{\lambda \omega}, 
{\cal M}_{8d}^{\mu \nu}= 
{\rm \hat {T_{8}}} {\cal M}_{2d}^{\lambda \omega},
\eer
where ${\rm \hat {T_{8}}}$ represents the replacement,   
$p_1 \leftrightarrow -p_3, p_2 \leftrightarrow -p_4, 
\mu \leftrightarrow \lambda$, and $\nu \leftrightarrow \omega$. 

For process 3, $\pi D^* \rightarrow \pi D^*$, we have
\begin{eqnarray}
{\cal M}_{3a}^{\nu \omega}&=& \grpp \grdsds~ 
\left ( -\imath \epsilon_{ijk} \tau^k \right )_{\alpha \beta} 
\left ( \frac{1} {t-m_\rho^2} \right ) \nonumber \\
&\times &
\left [ (u-s) g^{\nu \omega} + 4 (p_1^\nu p_3^\omega -p_3^\nu p_1^\omega) 
+ p_2^\nu (p_1+p_3)^\omega + (p_1+p_3)^\nu p_4^\omega \right ], \nonumber \\
{\cal M}_{3b}^{\nu \omega}&=& \gpdds^2~ (\tau^j \tau^i)_{\alpha \beta}
(-p_1-p_3+p_4)^\nu \left ( \frac{1}{u-m_D^2} \right )
(2p_1-p_4)^\omega, \nonumber \\
{\cal M}_{3c}^{\nu \omega}&=& \gpdds^2~
(\tau^i \tau^j)_{\alpha \beta}
(2p_1+p_2)^\nu \left ( \frac{1}{s-m_D^2} \right )
(-p_1-p_2-p_3)^\omega, \nonumber \\
{\cal M}_{3d}^{\nu \omega}&=& \gppdsds~ 
\delta_{ij} \delta_{\alpha \beta} g^{\nu \omega}.
\end{eqnarray}

For process 5, $\rho D \rightarrow \rho D$, we have
\begin{eqnarray}
{\cal M}_{5a}^{\mu \lambda}&=& \grrr \grdd~ 
\left ( -\imath \epsilon_{ijk} \tau^k \right )_{\alpha \beta} 
\left ( \frac{1} {t-m_\rho^2} \right ) \nonumber \\
&\times &
\left [ (u-s) g^{\mu \lambda} + 4 (p_2^\mu p_4^\lambda -p_4^\mu p_2^\lambda) 
+ p_1^\mu (p_2+p_4)^\lambda + 3(p_2+p_4)^\mu p_3^\lambda \right ], \nonumber \\
{\cal M}_{5b}^{\mu \lambda}&=& \grdd^2~ (-\tau^j \tau^i)_{\alpha \beta}
(-p_1+2p_4)^\mu \left ( \frac{1}{u-m_D^2} \right )
(-p_1+p_2+p_4)^\lambda, \nonumber \\
{\cal M}_{5c}^{\mu \lambda}&=& \grdd^2~
(-\tau^i \tau^j)_{\alpha \beta}
(p_1+2p_2)^\mu \left ( \frac{1}{s-m_D^2} \right )
(p_1+p_2+p_4)^\lambda, \nonumber \\
{\cal M}_{5d}^{\mu \lambda}&=& \grrdd~ 
\delta_{ij} \delta_{\alpha \beta} g^{\mu \lambda}.
\end{eqnarray}

For process 7, $\rho D^* \rightarrow \rho D^*$, we have
\begin{eqnarray}
{\cal M}_{7a}^{\mu \nu \lambda \omega} 
&=& \grrr \grdsds~ 
\left ( \imath \epsilon_{ijk} \tau^k \right )_{\alpha \beta} 
\left ( \frac{1} {t-m_\rho^2} \right ) \nonumber \\
&\times&
\left [ (p_1+p_3)^\gamma g^{\mu \lambda} + 
(p_1-2p_3)^\mu g^{\gamma \lambda} + 
(-2p_1+p_3)^\lambda g^{\mu \gamma} \right ] g_{\gamma \ggp} \nonumber \\
& \times &
\left [ (-p_2-p_4)^\ggp g^{\nu \omega} + 
(-p_1+p_2+p_3)^\omega g^{\ggp \nu} + 
(p_1-p_3+p_4)^\nu g^{\ggp \omega} \right ] ,\nonumber \\
{\cal M}_{7b}^{\mu \nu \lambda \omega} &=& \grdsds^2~ 
\left ( \tau^j \tau^i \right )_{\alpha \beta} 
\left ( \frac{1} {u-m_{D^*}^2} \right ) \nonumber \\
&\times &
\left [ (p_1+p_4)^\gamma g^{\mu \omega} + 
(p_1-2p_4)^\mu g^{\gamma \omega} + 
(-2p_1+p_4)^\omega g^{\gamma \mu} \right ] 
\left [ g_{\gamma \ggp}-\frac{(p_1-p_4)_\gamma 
(p_1-p_4)_\ggp}{m_{D^*}^2} \right ] \nonumber \\
& \times &
\left [ (p_1+p_3-p_4)^\nu g^{\ggp \lambda}  
-(p_2+p_3)^\ggp g^{\nu \lambda} + 
(-p_1+p_2+p_4)^\lambda g^{\ggp \nu} \right ] 
,\nonumber \\
{\cal M}_{7c}^{\mu \nu \lambda \omega} &=& \grdsds^2~
\left ( \tau^i \tau^j \right )_{\alpha \beta} 
\left ( \frac{1} {s-m_{D^*}^2} \right ) \nonumber \\
&\times&
\left [ (2p_1+p_2)^\nu g^{\gamma \mu} - 
(p_1+2p_2)^\mu g^{\nu \gamma} + 
(-p_1+p_2)^\gamma g^{\mu \nu} \right ] 
\left [ g_{\gamma \ggp}-\frac{(p_1+p_2)_\gamma 
(p_1+p_2)_\ggp}{m_{D^*}^2} \right ] \nonumber \\
& \times &
\left [ (-p_3+p_4)^\ggp g^{\lambda \omega}  
+(-p_1-p_2-p_4)^\lambda g^{\ggp \omega} + 
(p_1+p_2+p_3)^\omega g^{\ggp \lambda} \right ] 
,\nonumber \\
{\cal M}_{7d}^{\mu \nu \lambda \omega}&=& \grrdsds~ \left [ 
(2\tau^j \tau^i-\tau^i \tau^j)_{\alpha \beta} g^{\mu \nu} g^{\lambda \omega}
+(2\tau^i \tau^j-\tau^j \tau^i)_{\alpha \beta} g^{\mu \omega} g^{\nu \lambda}
-2 \delta_{ij} \delta_{\alpha \beta} g^{\mu \lambda} g^{\nu \omega} \right ].
\label{mf}
\end{eqnarray}

After averaging (summing) over initial (final) spins and isopsins,
the cross section for a process is given by 
\begin{eqnarray}
\frac {d\sigma_n}{dt}&=& \frac {1}{64 \pi s p_{i,\rm cm}^2 I_s I_i}
{\cal M}_n^{\lambda_k \cdots \lambda_l} 
{\cal M}_n^{* \lambda_k^\prime \cdots \lambda_l^\prime}
\left ( g_{\lambda_k \lambda_k^\prime}-\frac{p_{k \lambda_k} 
p_{k \lambda_k^\prime}} {m_k^2} \right )
\cdots 
\left ( g_{\lambda_l \lambda_l^\prime}-\frac{p_{l \lambda_l} 
p_{l \lambda_l^\prime}} {m_l^2} \right ), 
\end{eqnarray}
with $s,t,u$ being the standard Mandelstam variables, and 
\begin{eqnarray}
p_{i,\rm cm}^2=\frac {\left [ s-(m_1+m_2)^2 \right ]
\left [ s-(m_1-m_2)^2 \right ]}{4s}
\end{eqnarray}
is the squared momentum of initial-state mesons in the 
center-of-momentum frame. 
The factors $I_s$ and $I_i$ are due to averaging over initial spin and 
isospins, respectively. Values of $I_s$ are 1, 1, 3, 3, 3, 3, 9 and 9, 
respectively, for processes $1$ to $8$ in Fig. \ref{diagrams}, 
while $I_i$ is 6 for all processes.

%%%%%%%%%%%%%%%%%%%%
\section*{Appendix B}
\setcounter{equation}{0}
\def\theequation{B\arabic{equation}}

In this appendix, we show as an example that 
the scattering amplitude for process 2, $\pi D\to\rho D^*$,
satisfies the condition of Eq. (\ref{conservation}) as a result of
the SU(4) flavor current conservation. In particular, 
we shall prove that 
${\cal M}_2^{\lambda \omega} p_{4 \omega}=0$.
Starting from Eq. (\ref{m2}), we obtain
\ber
{\cal M}_{2a}^{\lambda \omega} p_{4 \omega} &=& \grpp \gpdds~ 
\left ( \imath \epsilon_{ijk} \tau^k \right )_{\alpha \beta} 
\left ( \frac {t-m_D^2}{t-m_\pi^2} \right ) (-2 p_1)^\lambda, \nonumber \\
{\cal M}_{2b}^{\lambda \omega}p_{4 \omega} &=& \gpdds \grdd~
\left ( \tau^j \tau^i \right )_{\alpha \beta} 
\left ( \frac {u-m_\pi^2}{u-m_D^2} \right ) 
(-p_1+p_2+p_4)^\lambda, \nonumber \\
{\cal M}_{2c}^{\lambda \omega}p_{4 \omega} &=& \gpdds \grdsds~
\left ( \tau^i \tau^j \right )_{\alpha \beta} 
\left ( \frac {1}{s-m_{D^*}^2} \right )
\left [ \left (s-m_\rho^2 \right ) (p_1-p_2)^\lambda \nonumber \right . \\
&+& \left . \frac {(m_D^2-m_\pi^2)(m_{D^*}^2-m_\rho^2)} {m_{D^*}^2} 
p_4^\lambda \right ] ,\nonumber \\
{\cal M}_{2d}^{\lambda \omega}p_{4 \omega}&=& \gprdds~ 
\left ( \tau^i \tau^j -2 \tau^j \tau^i \right )_{\alpha \beta} p_4^\lambda.
\eer
In arriving at the above, 
we have discarded all terms with $p_3^\lambda$ as they vanish 
after contracting with the polarization vector $\epsilon_{3\lambda}$. 
Using the SU(4) relation for the coupling constants 
shown in Eq. (\ref{coupling}) then gives 
\ber
{\cal M}_2^{\lambda \omega} p_{4 \omega}
&=& 
\frac{g^2}{8} \left (\tau^j \tau^i \right )_{\alpha \beta} 
\left (m_D^2-m_\pi^2 \right ) 
\left ( 
\frac {-p_1^\lambda}{t-m_\pi^2} +\frac {p_2^\lambda}{u-m_D^2} \right )
\nonumber \\
&+& 
\frac{g^2}{16} \left (\tau^i \tau^j \right )_{\alpha \beta} 
\left [ 2\left (\frac {m_D^2-m_\pi^2}{t-m_\pi^2}\right ) p_1^\lambda 
+\left (\frac {m_{D^*}^2-m_\rho^2}{s-m_{D^*}^2} \right ) 
(p_1-p_2)^\lambda \right .\nonumber \\
&+& \left . \frac {(m_D^2-m_\pi^2)(m_{D^*}^2-m_\rho^2)}
{m_{D^*}^2 (s-m_{D^*}^2)} (p_1+p_2)^\lambda \right ] .
\eer
With degenerate pseudoscalar meson masses and degenerate vector meson masses,  
the above expression then reduces to zero. 
For other amplitudes shown in Eqs. (\ref{mi})-(\ref{mf}), 
the current conservation condition can be similarly proved. 
We note that ${\cal M}_2^{\lambda \omega} p_{3 \lambda}=0$ holds for 
any masses.

%%%%%-------------------------References---------------------------------
\pagebreak
{}


\begin{thebibliography}{99}
\bibitem{err}
We have realized that there is a typo in the expression for 
the thermal averaged cross section in Ref. \cite{ds}. 
The factors $\alpha_1$ and $\alpha_2$ 
in the denominator outside the Bessel functions 
should be $\alpha_1^2$ and $\alpha_2^2$, respectively.
\end{thebibliography}
\end{document}